\documentclass{article}

\usepackage{spconf,amsmath,graphicx,url,microtype}
\usepackage{tikz}
\usepackage{textcomp}
\usepackage{lipsum}
\usepackage{mathtools}

\newcommand\copyrighttext{%
  \footnotesize \textcopyright 2022 IEEE. Personal use of this material is permitted.
  Permission from IEEE must be obtained for all other uses, in any current or future 
  media, including reprinting/republishing this material for advertising or promotional 
  purposes, creating new collective works, for resale or redistribution to servers or 
  lists, or reuse of any copyrighted component of this work in other works. }
\newcommand\mycopyrightnotice{%
\begin{tikzpicture}[remember picture,overlay]
\node[anchor=south,yshift=10pt] at (current page.south) {\fbox{\parbox{\dimexpr\textwidth-\fboxsep-\fboxrule\relax}{\copyrighttext}}};
\end{tikzpicture}%
}

\DeclareMathOperator*{\argmax}{argmax}

\title{Improving Lyrics Alignment through Joint Pitch Detection}
\name{Jiawen Huang \sthanks{JH is a research student at the UKRI Centre for Doctoral Training in Artificial Intelligence and Music, supported jointly by UK Research and Innovation [grant number EP/S022694/1] and Queen Mary University of London.}\textsuperscript{1}, Emmanouil Benetos\textsuperscript{1}, Sebastian Ewert\textsuperscript{2}}
\address{\textsuperscript{\rm 1} Centre for Digital Music, Queen Mary University of London, UK \\
\textsuperscript{\rm 2} Spotify}
\begin{document}
\ninept
\maketitle

\begin{abstract}

In recent years, the accuracy of automatic lyrics alignment methods has increased considerably. Yet, many current approaches employ frameworks designed for automatic speech recognition (ASR) and do not exploit properties specific to music. Pitch is one important musical attribute of singing voice but it is often ignored by current systems as the lyrics content is considered independent of the pitch.
In practice, however, there is a temporal correlation between the two as note starts often correlate with phoneme starts.
At the same time the pitch is usually annotated with high temporal accuracy in ground truth data while the timing of lyrics is often only available at the line (or word) level. In this paper, we propose a multi-task learning approach for lyrics alignment that incorporates pitch and thus can make use of a new source of highly accurate temporal information. Our results show that the accuracy of the alignment result is indeed improved by our approach. 
As an additional contribution, we show that integrating boundary detection in the forced-alignment algorithm reduces cross-line errors, which improves the accuracy even further. 

\end{abstract}
\mycopyrightnotice

\section{Introduction}

Given the lyrics to a song, audio-to-lyrics alignment aims at identifying for each lyric the corresponding position in a recording of the song, at a line, word or phoneme level \cite{fujihara2012lyrics}.
Applications include the generation of karaoke-style lyrics for music players and subtitles for music videos. Moreover, lyrics alignment can serve as a basic block for singing voice analysis and can benefit other tasks such as cover song identification and music structure analysis.

Since the release of large datasets providing audio with lyrics annotations, such as DALI \cite{meseguer2020creating} and DAMP \cite{damp, gupta2018semi}, there has been significant progress in reducing the alignment error. 
Many previous methods are adapted from automatic speech recognition (ASR) \cite{gupta2018semi, gupta2019acoustic, stoller2019end, sharma2019automatic}.
However, singing voice tends to be more complex than speech signals. Singing voice has a wider dynamical range, the pronunciation of words varies a lot more, while singing techniques allow the artist to control the sound in various ways \cite{fujihara2012lyrics, kruspe2014keyword, kruspe2016bootstrapping, loscos1999low}. Moreover, the background music is often highly correlated with the singing voice, making it more challenging to analyze the vocals with high accuracy. Gupta et al.~\cite{gupta2019acoustic} explored music-related features and adapted the acoustic model on polyphonic data.
In their follow-up work \cite{gupta2020automatic}, they trained a genre-informed acoustic model and found improvements in both lyrics alignment and transcription. These approaches indicate that domain knowledge could help train the acoustic model.

One challenge of training the acoustic model is the lack of fine-grained annotation. A frame-level annotation of lyrics is hard to obtain. Some previous works trained a Gaussian mixture hidden Markov model (GMM-HMM) to predict the frame-level annotation and take it as the ground truth for training \cite{demirel2020automatic, gupta2019acoustic}. Others adopted the connectionist temporal classification (CTC) loss \cite{graves2006connectionist} and trained the acoustic model in an end-to-end way \cite{stoller2019end, vaglio2020multilingual}.
However, the CTC loss is a weaker form of supervision as it only enforces that a symbol is observed but does not specify when \cite{liu2018connectionist}.


In this paper, we explore multi-task learning \cite{DBLP:journals/ml/Caruana97} for lyrics alignment, where the auxiliary task is pitch detection. At first this seems to be an unusual combination as in traditional speech synthesis, pitch and lyrics are modeled as two independent attributes and thus it might not be possible to share many feature representations between the two tasks - an aspect that typically drives which tasks in multitask learning are combined. However, using this unusual combination we can integrate pitch as an additional source of highly accurate temporal information. In particular, the pitch output of our method has to be temporally precise to achieve high accuracy and the idea is to evaluate whether this might act as an inductive bias that encourages the lyrics output of the network to be more precise as well. In other words, by adding a frame-level loss function through multi-task learning, it implicitly applies a stronger restriction about the timing as many of the phonemes and pitches share onsets. Additionally, previous studies have shown that joint learning of unrelated tasks might lead to sparser and more informative representations \cite{paredes2012exploiting}, which could lead to improved accuracy as well. 
Hung et al. \cite{hung2019multitask} proposed a multi-task learning method for instrument detection which models pitch and instrument.
Inspired by their work, we propose a model that outputs a representation for both pitch~\footnote{In this study, pitch detection refers to the quantized pitch (as typically specified as the pitch of a note in MIDI) as opposed to the fundamental frequency.} and phoneme, where the task-wise loss can be computed by applying pooling along the other axis. We favor this method over a multi-head architecture
commonly used in multi-task learning as the pitch information is kept close to the phoneme information all the way.


In real-world applications such as generating karaoke-style scrolling lyrics, cross-line misalignment is less tolerable than inline errors, and should therefore be avoided. Just as lyrics can be segmented line by line, the corresponding cut points in audio can be regarded as line-level boundaries, which can be estimated by a network. Previous works in audio-to-score alignment show that onset information can enhance the temporal precision by adding an onset term to the cost function \cite{EwertMG09_HighResAudioSync_ICASSP}. Similarly, we propose to incorporate the boundary probability of audio frames into the alignment algorithm to further improve the alignment performance.

To the best of the authors' knowledge, this is the first attempt to apply multi-task learning to lyrics alignment. Although multi-task learning has been extensively explored in ASR \cite{yang2018joint, kim2017joint, xiao2018hybrid}, we choose a musically meaningful auxiliary task to take advantage of the correlation between lyrics and pitch \cite{nichols2016relationships, gong2015real}. This is also the first time an audio-to-lyrics alignment system incorporates boundary information into the forced-alignment.

The paper is structured as follows: The proposed architecture and loss computations are discussed in Sec.~\ref{sec:acoustic}. The proposed method to incorporate boundary information is explained in Sec.~\ref{sec:align}. The dataset, baseline and experimental settings are described in Sec.~\ref{sec:exp}. The results are presented and discussed in Sec.~\ref{sec:res}. Finally, we conclude and look into future directions in Sec.~\ref{sec:conclusion}.

\section{Joint Phoneme Recognition and Pitch Detection}\label{sec:acoustic}

\begin{figure}[t]
  \centering
  \includegraphics[width =\columnwidth]{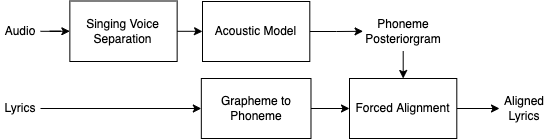}
  \caption{System overview.}
  \label{fig:diagram}
\end{figure}

A general overview of the various modules involved in our system is given in Fig.~\ref{fig:diagram} -- our proposed extensions and their details will be discussed below. It consists of a singing voice separation module, an acoustic model, a grapheme to phoneme (g2p) module, and the alignment algorithm. The lyrics are first converted to a phoneme sequence via g2p (Sec.~\ref{sec:dataset}). The singing voice separation model extracts the vocals to remove the effect of the background music (Sec.~\ref{sec:dataset}). The acoustic model takes the Mel-spectrogram of the separated vocals as input, and produces the phoneme posteriorgram (Sec.~\ref{sec:ac_model}). Then the forced-alignment algorithm is applied on the posteriorgram and the corresponding lyrics (Sec.~\ref{sec:align}).

\subsection{Acoustic Model}\label{sec:ac_model}


The acoustic model takes a 128-bin Mel-spectrogram with a fixed duration as input. The sampling rate is 22050 Hz, the fft size is 512 and the hop size is 256. 
To fix notation, let $X = \{x_1, x_2, ...x_m\}$ be the frames of the input Mel-spectrogram, and $L = \{l_1, l_2,...l_n\}$ be the phoneme sequence for the corresponding lyrics, where $m$ and $n$ are the lengths of the Mel-spectrogram and the phoneme sequence. The size of the phoneme set $\mathcal{S}_{\text{phone}}$ is $N_{\text{phone}}$, and the size of the pitch set $\mathcal{S}_{\text{pitch}}$ is $N_{\text{pitch}}$. 

\begin{figure}[t]
  \centering
  \includegraphics[width =0.95\columnwidth]{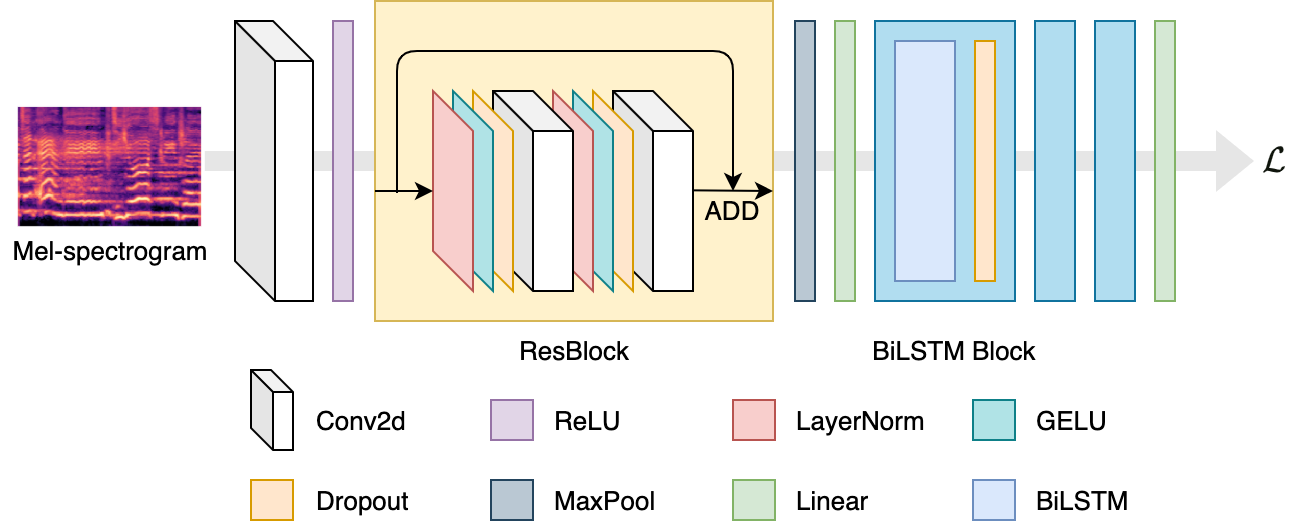}
  \caption{The network architecture for the baseline, boundary detection, and the MTL models. The only differences among them are the dimensions of the BiLSTMs and the last fully-connected layer.}
  \label{fig:baseline}
\end{figure}

Fig. \ref{fig:baseline} shows the network architecture.
It consists of a convolutional layer, a residual convolutional block, a fully-connected layer, 3 bidirectional LSTM (Long Short-Term Memory) layers, a final fully-connected layer, and non-linearities in between. The kernel size of the convolutional layers is $3 \times 3$, with stride and padding equal to one. The number of filters for the three convolutional layers is 32. The dimensions of the bidirectional LSTMs (BiLSTM) are 256. Layer normalization is applied on the feature dimension, with a mini-batch size of 128. All dropout rates are set to 0.1. The last fully-connected layer is time-distributed (applied to each frame), with a target size of $N_{\text{phone}} \times N_{\text{pitch}}$.
After that, the output is reshaped to an order-3 tensor $\mathcal{D}$ of size $N_{\text{time}} \times N_{\text{phone}} \times N_{\text{pitch}}$, where $N_{\text{time}}$ is the output frame number. 
This representation $\mathcal{D}$ can be considered as the joint probability distribution before a \texttt{softmax} operation.
This model is later referred to as \textbf{MTL} (\textbf{M}ulti-\textbf{T}ask \textbf{L}earning).

\subsection{Loss Function}\label{sec:loss}

\begin{figure*}[ht!]
  \centering
  \includegraphics[width =1.7\columnwidth]{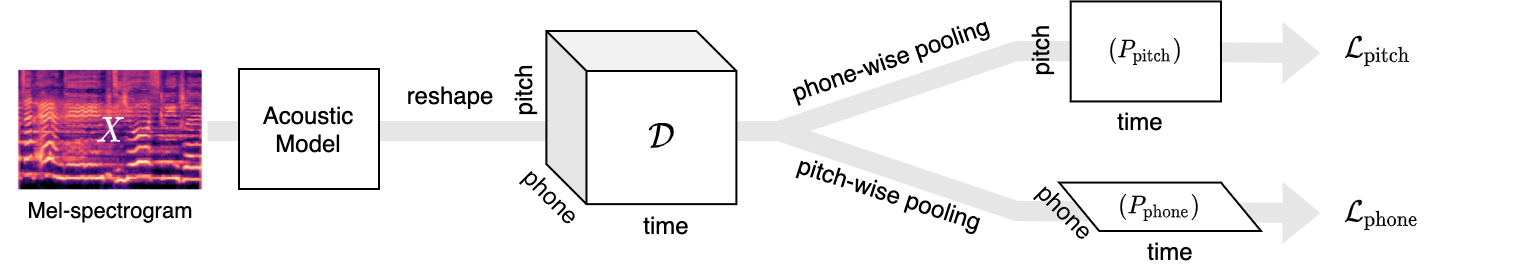}
  \caption{Multi-task loss computation.}
  \label{fig:pitch_mtl}
\end{figure*}

Fig. \ref{fig:pitch_mtl} describes the steps to compute the loss functions. Average pooling is applied on $\mathcal{D}$ along the pitch axis to get the (log-) posteriorgram of phoneme $P_{\text{phone}}$ of size $N_{\text{time}} \times N_{\text{phone}}$, where the CTC loss \cite{graves2006connectionist} $\mathcal{L}_{\text{phone}}$ is computed. The same is applied along the phoneme axis to get the pitch posteriorgram $P_{\text{pitch}}$ of size $N_{\text{time}} \times N_{\text{pitch}}$ and compute the frame-level cross entropy loss $\mathcal{L}_{\text{pitch}}$. The final loss is a weighted sum of these two:
\begin{equation}
 \mathcal{L}  =  \mathcal{L}_{\text{phone}} + \lambda \cdot  \mathcal{L}_{\text{pitch}} \label{eq:1}
\end{equation}
where $\lambda$ is an adjustable parameter. We tested different values in $\{0.5, 0.8, 1.0, 1.2, 1.5\}$ and found $\lambda = 0.5$ to yield the best results and thus we used this setting in our experiments below.

The baseline model shares the same architecture except for the dimension of the last fully-connected layer, but focuses on phoneme recognition only. The output is a $N_{\text{time}} \times N_{\text{phone}}$ posteriorgram, where the CTC loss is computed.

\section{Alignment}\label{sec:align}

\subsection{Viterbi Forced Alignment}\label{sec:std_align}

In the calculation of the CTC loss \cite{graves2006connectionist}, the probabilities of all possible alignment paths are accumulated. These paths are generated by inserting blank ($\epsilon$) and repeated labels to the original sequence. By applying the same rules to the lyrics to be aligned at inference time, the best path can be decoded via Viterbi forced alignment \cite{forney1973viterbi}.

To be more specific, the phoneme sequence is expanded by inserting $\epsilon$ symbols in between the phonemes to $L^{*} = \{\epsilon, l_1, \epsilon, l_2,...\epsilon, l_n, \epsilon\}$. Let $P_{\text{phone}}(t, p)$ be the log-probability of the input frame $x_t$ being a phoneme $p \in \mathcal{S}_{\text{phone}}$. A path $\{(t_q, l^{*}_q)\}$ is a sequence of index pairs indicating $x_{t_q}$ is aligned to $l^{*}_q$.
The target is to find the path that maximizes the score:
\begin{equation}
\argmax_{path} \sum_{(t, l^{*}) \in path} P_{\text{phone}}(t, l^{*})\label{eq:2}
\end{equation}
The solution can be computed efficiently via dynamic programming.

\subsection{Incorporating Boundary Information}\label{sec:bdr}

A boundary detection model is trained independently to predict the line-level boundary probability. The input and the network architecture are the same as the baseline except that it has a smaller BiLSTM dimension 32 and output size is 1 per time step. The target label is a boundary activation curve (range: 0 $\sim$ 1). The start time of a line of lyrics is considered a boundary event. Each event is converted to a Gaussian window centered at the event time, at a size of 0.7~sec.
Let $P_{\text{bdr}}(t)$ be the boundary log-probability of $x_t$ predicted by the model.

We propose to add the boundary probability to the Viterbi score as a bonus at the line beginning of the lyrics, which is a `boundary' in the text. This is to encourage aligning a phoneme at the beginning of a line to a boundary-like audio frame.
Equation \eqref{eq:2} is updated as follows:
\begin{equation}
\argmax_{path} \sum_{(t, l^{\star}) \in path} P_{\text{phone}}(t, l^{\star}) + \alpha \sum_{\substack{(t, l^{\star}) \in path \\ l^{\star} \text{ is a line start}}} P_{\text{bdr}}(t) \label{eq:3}
\end{equation}
where $\alpha$ is an adjustable weighting parameter. We test different values in $\{0.5, 0.8, 1.0, 1.2, 1.5\}$ and found $\alpha = 0.8$ to yield the best results and thus we used this setting in our experiments below. This alignment method is referred to as \textbf{BDR} (\textbf{B}oun\textbf{D}a\textbf{R}y) later in text.

\begin{table}[t!]
\centering
\begin{tabular}{|l|c|c|l|l|l|l|}
\hline
Dataset                             & SS & FW    & \multicolumn{2}{l|}{Jamendo} & \multicolumn{2}{l|}{Mauch} \\ \hline
Metric                              & -  &  -           & AAE               & PCO               & AAE               & PCO           \\ \hline \hline
SDE2 \cite{stoller2019end}          & Y  &  E2E         & 0.39              & 0.87              & 0.26              & 0.87         \\ \hline
GC \cite{gupta2020automatic}        & N  &  STD         & 0.22     &    0.94  & 0.19     & 0.91           \\ \hline
VHM \cite{vaglio2020multilingual}   & Y  &  E2E         & 0.37              & 0.92              & 0.22              & 0.91  \\ \hline \hline
Baseline                            & Y  &  E2E         & 0.31              & 0.94              & 0.20              & 0.89             \\ \hline
\end{tabular}
\caption{Comparison with the state-of-the-art systems for lyrics alignment. The \textbf{SS} column indicates if the input has been source-separated (Yes or No). The \textbf{FW} column indicates if the method takes an end-to-end (E2E) or standard (STD) ASR framework.}
\label{tab:sota}
\end{table}

\section{Experiments}\label{sec:exp}

\subsection{Dataset}\label{sec:dataset}
All models are trained on the DALI v2 dataset \cite{meseguer2020creating}. There are 7756 songs in total with word-level lyrics annotations and note-level pitch annotations. We only use the English subset according to the language label. The training set contains 4224 songs and the validation set contains 1056 songs for the baseline and the multi-task model~\footnote{The data splits and the code can be accessed through the link: \url{https://github.com/jhuang448/LyricsAlignment-MTL}}. The samples are generated by applying a 5.6~sec sliding window with a hop size of 2.8~sec. The target lyrics for an audio segment are the words fully covered within the window.

For the boundary detection model, a subset of the above validation set is left out for evaluation. The subset contains 45 songs.

For lyrics alignment, we run evaluations on the Jamendo \cite{ramona2008vocal, stoller2019end} and Mauch \cite{mauch2011integrating} datasets. Each of them has 20 Western pop songs and word-level timestamps and boundary annotations. They are also used in the MIREX~\footnote{\url{https://www.music-ir.org/mirex/}} lyrics alignment challenges.

For pitch detection, we evaluate the MTL model using the RWC Music Database - Popular Music \cite{goto2006aist}. It contains 94 popular songs (74 Japanese and 20 English)~\footnote{The dataset contains 100 songs, but 6 of them (No. 3, 5, 8, 10, 23, and 66) are removed because they have multiple singers.}.

Vocals are extracted by an implementation of \cite{jansson2017singing} for all  songs mentioned above before feeding to the network.
The open-source g2p tool~\footnote{\url{https://github.com/Kyubyong/g2p}} is used to convert the lyrics to phoneme sequences. The phoneme set follows the convention of the CMU pronouncing dictionary~\footnote{\url{http://www.speech.cs.cmu.edu/cgi-bin/cmudict}} and has a size of 39. In practice, we added the space~\textvisiblespace \, and the epsilon $\epsilon$ (for the CTC loss) to the phoneme set so that the number of classes $N_{\text{phone}}$ is 41.
The target pitch range is D2-C6, therefore $N_{\text{pitch}}$ is 47 (with one additional class for silence).

\subsection{Training}\label{training}

The models are trained using a learning rate of $1e^{-4}$ and the ADAM optimizer \cite{kingma2014adam}. Early stopping is adopted when the validation loss does not decrease for 20 epochs after the 20$^{th}$ epoch. We employed this strategy as we found some early checkpoints to produce a relatively low loss value but which turned out to be underfitting. The boundary detection model is trained with binary cross entropy loss, and we do not wait for 20 epochs before early-stopping.


\begin{table}[t]
\centering
\begin{tabular}{|l|l|l|l|}
\hline
Metric        & COnPOff      & COnP      & COn      \\ \hline \hline
pYIN \cite{mauch2014pyin} & 5.9\%             & 12.2\%          & 43.5\%         \\ \hline
MTL                       & 4.9\%            &   16.3\%         &  32.6\%        \\ \hline 
\end{tabular}
\caption{Pitch detection results.}
\label{tab:pitch}
\end{table}

\begin{table}[t]
\centering
\begin{tabular}{|l|l|l|l|l|}
\hline
Metric &         Precision           &     Recall        &     F-score  & AUC \\ \hline \hline
BDR   &          79.3\%            &     55.9\%               &    64.3\%  & 89.9\%             \\ \hline
\end{tabular}
\caption{Boundary detection results.}
\label{tab:bdr}
\end{table}

\begin{table*}[ht!]
\centering
\begin{tabular}{|l|l|l|l|l|l|l|l|l|}
\hline
Level      & \multicolumn{4}{l|}{Word} & \multicolumn{4}{l|}{Line} \\ \hline
Dataset          & \multicolumn{2}{l|}{Jamendo} & \multicolumn{2}{l|}{Mauch} & \multicolumn{2}{l|}{Jamendo} & \multicolumn{2}{l|}{Mauch} \\ \hline
Metric           & AAE      & PCO     & AAE     & PCO       & AAE      & PCO     & AAE     & PCO    \\ \hline \hline
Baseline         & 0.31     & 0.94    & 0.20    & 0.89      & 0.30     & 0.93    & 0.37    & 0.87   \\ \hline
Baseline + BDR   & 0.29     & 0.94    & 0.20    & 0.90      & 0.25     & 0.95    & 0.33    & 0.88   \\ \hline
MTL              & 0.23     & 0.94    & 0.21    & 0.90      & 0.27     & 0.93    & 0.43    & 0.86   \\ \hline
MTL + BDR        & 0.23     & 0.94    & 0.20    & 0.91      & 0.25     & 0.94    & 0.42    & 0.89   \\ \hline
\end{tabular}
\caption{Lyrics alignment results. Significant improvement in AAE on Jamendo for the MTL model can be observed for word-level metrics. All line-level metrics increase after introducing boundary information.}
\label{tab:mtl}
\end{table*}

\subsection{Evaluation}\label{sec:eval}


For lyrics alignment at the word level, we report the Average Absolute Error (AAE) \cite{mesaros2008automatic} and Percentage of correct onsets with a tolerance window of 0.3 seconds (PCO) \cite{mauch2011integrating}. These metrics are averaged over all the songs in the dataset. In addition to the word-level evaluation, we also report the line-level results to demonstrate the benefit from boundary information.

For boundary detection, we compute the precision, recall, and F-measure. A predicted boundary is considered a hit if there is a reference boundary within $\pm$0.5 seconds \cite{turnbull2007supervised}. We also report the Area Under the Receiver Operating Characteristic Curve (AUC-ROC) between the boundary activation curve and the ground truth.

For pitch detection, we compute 3 F-scores (COn, COnP, and COnPOff) originally proposed in \cite{molina2014evaluation}. COn only evaluates the onset times, COnP evaluates both onsets and the pitches, and COnPOff evaluates onsets, pitches, and offsets. The onset tolerance is 50ms, while the offset tolerance is max(50ms, 0.2*note duration). Since the ground truth of RWC Pop contains many octave errors, instead of evaluating the absolute pitch, we compute the metrics on a octave-wrapped pitch with a tolerance of 50 cents.

\section{Results and Discussion}\label{sec:res}
\subsection{State of the Art Comparison}

The comparison of our baseline with state-of-the-art lyrics alignment systems is listed in Tab.~\ref{tab:sota}. Both GC \cite{gupta2020automatic} and VHM \cite{vaglio2020multilingual} are trained on DALI v1, while VHM takes an end-to-end approach on separated vocals and GC uses a standard ASR framework trained on polyphonic audio. SDE2 \cite{stoller2019end} is a wave-U-net model trained on an internal dataset consisting 44,232 songs. It is also trained on separated vocals.

The results show that our baseline performs similarly to the current state-of-the-art on both Jamendo and Mauch. Among the E2E models trained on separated vocals, our baseline outperforms SDE2 and VHM on all metrics except PCO on Mauch. Besides the network architecture, the key difference is the alignment unit. In our baseline, the CTC loss and the alignment are computed on phoneme sequences, while SDE2 and VHM are computed on characters. This further 
verifies that using phonemes over characters for alignment has better performance \cite{vaglio2020multilingual}.


\subsection{Boundary Detection and Pitch Detection}
The boundary detection results are listed in Tab.~\ref{tab:bdr}. An AUC of 89.9\% indicates it is a decent boundary detector. However, it is hard to provide comparison with existing systems because most of them detect structural boundaries (segmentation).


The pitch detection results of the MTL model are listed in Tab.~\ref{tab:pitch}. For comparison, we also list the pYIN \cite{mauch2014pyin} (note mode) results. Our model is trained on DALI, which is mostly Western music, while 80\% of the songs in the testing set are in Japanese. We argue that there is a large difference in data distribution between the training and testing sets, leading to a limited performance. 

\begin{figure}[ht!]
  \centering
  \includegraphics[width =\columnwidth]{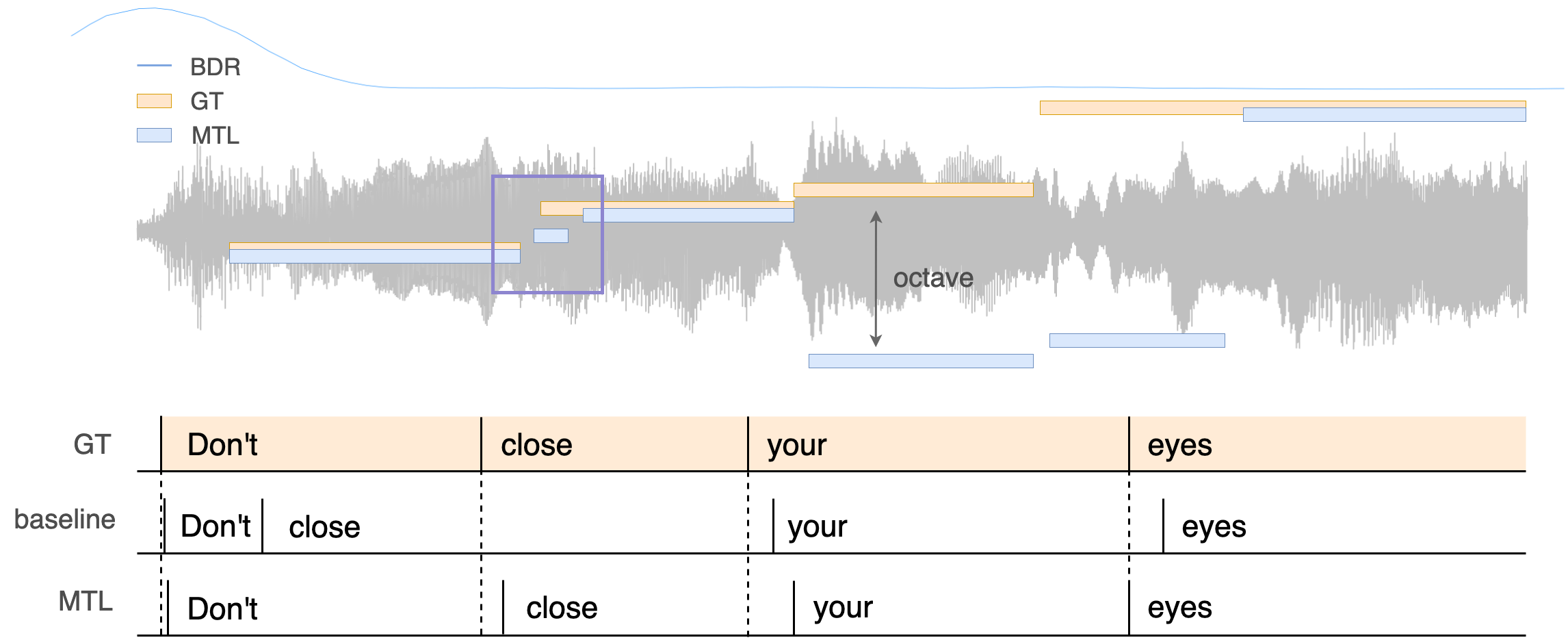}
  \caption{Lyrics alignment, pitch detection, and boundary detection results on a clip of \texttt{Avercage\_-\_Embers.mp3} from Jamendo. The GT (ground truth) notes are slightly shifted for visualization purposes. Two notes have the same pitch if there is an overlap.}
  \label{fig:exp}
\end{figure}

The predicted pitch results and the ground truth on a short clip from Jamendo are visualized in Fig.~\ref{fig:exp}. It can be observed that our pitch tracker generally follows the singing voice, but produces some short notes that should be merged.

\subsection{MTL Evaluations}

Results of different combinations of Baseline / MTL with BDR are listed in Tab.~\ref{tab:mtl}.
Comparing with the baseline, the improvement is significant in AAE on Jamendo for the MTL model, but limited in the other metrics. As can be observed in Tab.~\ref{tab:sota}, results on Jamendo are worse than those on Mauch. This is probably due to more slurred pronunciation in Jamendo \cite{stoller2019end}. We argue that our MTL model alleviates this problem by estimating the pitch at the same time. Since pitch and phoneme often share onsets, a change in pitch can be a good indicator of a change in phoneme as well. The lyrics alignment results on the same clip are presented in Fig.~\ref{fig:exp}. Comparing to the baseline, MTL is able to place the word ``close" at the right time with the help of changes detected in pitch (within the purple frame).

Though the benefit from adding boundary information is marginal in word-level metrics, it is clear in line-level. Both metrics on Jamendo and Mauch are better with BDR.

\section{Conclusion}\label{sec:conclusion}

In this work, we propose a multi-task learning approach for lyrics-to-audio alignment by learning a joint representation for pitch and phoneme, and add boundary information to enhance the alignment. The proposed approach is built upon a phoneme-based end-to-end acoustic model as the baseline, which outperforms state-of-the-art end-to-end systems for lyrics alignment.  The proposed model also outperforms the baseline in both word-level and line-level metrics.


Meanwhile, we recognize some limitations of our system.
By introducing the additional boundary model, the performance is slightly improved at the cost of efficiency ($\times 1.3$ computation time). Besides, the pitch concept might not apply well to speech and rap in music. One possible solution is to add one pitch class for such voice.
For future work, we plan to evaluate the acoustic model on lyrics transcription as well. To further extend the application scenario, we plan to align the pitched notes to the lyrics with the learned representation $\mathcal{D}$. In this way, we can achieve melody and lyrics transcription of singing voice in one pass.

\bibliographystyle{IEEEbib}
\bibliography{biblio}

\end{document}